\documentclass[aps,pre,twocolumn,groupedaddress,floatfix,longbibliography,superscriptaddress]{revtex4-2}

\usepackage{amsmath}
\usepackage{amssymb}
\usepackage{amsfonts}
\usepackage{graphicx}
\usepackage{bbold}
\usepackage{bm}
\usepackage{bm}
\usepackage{cancel}
\usepackage{times,float}
\usepackage{graphicx}
\usepackage[usenames,dvipsnames,svgnames]{xcolor}
\usepackage{hyperref}
\hypersetup{colorlinks=true, linkcolor=Blue, citecolor=Blue,urlcolor=Blue}
\usepackage{multirow}
\usepackage{ulem}
\usepackage{color,epsfig}
\usepackage{bm}
\usepackage{slashed}
\usepackage{hyperref}
\usepackage{subfigure}
\usepackage{feynmf}
\usepackage{array}
\usepackage{orcidlink}

\newcommand{\bea}{\begin{eqnarray}}
\newcommand{\eea}{\end{eqnarray}}
\newcommand{\be}{\begin{equation}}
\newcommand{\ee}{\end{equation}}

\newcommand{\nn}{\nonumber}
\newcommand{\ii}{\mathrm{i}}

\newcolumntype{P}[1]{>{\centering\arraybackslash}p{#1}}

\begin{document}

\title{Topological versus conventional superconductivity in a Weyl semimetal: A microscopic approach}
\author{Enrique Mu\~noz~\orcidlink{0000-0003-4457-0817} }
\email{ejmunozt@uc.cl}
\affiliation{Facultad de F\'isica, Pontificia Universidad Cat\'olica de Chile, Vicu\~{n}a Mackenna 4860, Santiago, Chile}
\affiliation{Center for Nanotechnology and Advanced Materials CIEN-UC, Avenida Vicuña Mackenna 4860, Santiago, Chile}

\author{Juan Esparza}

\affiliation{Facultad de F\'isica, Pontificia Universidad Cat\'olica de Chile, Vicu\~{n}a Mackenna 4860, Santiago, Chile}
\author{Jos\'e Braun}

\affiliation{Facultad de F\'isica, Pontificia Universidad Cat\'olica de Chile, Vicu\~{n}a Mackenna 4860, Santiago, Chile}
\author{Rodrigo Soto-Garrido~\orcidlink{0000-0002-6593-5443} }
\email{rodsoto@uc.cl}
\affiliation{Facultad de F\'isica, Pontificia Universidad Cat\'olica de Chile, Vicu\~{n}a Mackenna 4860, Santiago, Chile}

\begin{abstract}
Starting from a microscopic model for the particle-particle interactions in a Weyl semimetal, we analyzed the possibility for conventional as well as monopole Cooper pairing between quasiparticle excitations at the same (intra-nodal) or opposite (inter-nodal) Weyl nodes. We derived a coupled system of self-consistent BCS-like equations, where the angular dependence of the pairings is directly determined from the microscopic interaction symmetries. We studied the competition between conventional and monopole superconducting phases, thus obtaining explicitly the phase diagrams from the microscopic interaction model parameters. We determined the critical temperatures for both phases, and the low temperature critical behavior, including the specific heat, that we suggest as possible experimental probe for topological quantum criticality in Weyl semimetals.
\end{abstract}

\maketitle 

\section{Introduction}

Topological semimetals are characterized by the presence of nodal points connecting the valence and conduction bands, thus leading to a variety of emerging low-energy quasiparticle excitations~\cite{Wehling_014,Chiu_016,Bradlyn_016,Armitage_018}. In Weyl semimetals (WSMs), due to the presence of pairs of gapless nodal points with nearly linear dispersion, those quasiparticles are pseudorelativistic Weyl fermions~\cite{Burkov_018,Armitage_018}, experimentally detected in compounds from the transition metal monopnictide family (TaAs, TaP, NbAs and NbP)~\cite{Xu_015,Grassano_018,Zhang_016,Arnold_016,Shekhar_015}. Each Weyl node is a monopolar source of Berry curvature, and hence they are protected from being gapped since their monopole charge $\mathcal{C} = \pm 1$ (chirality) is a topological invariant~\cite{Burkov_018,Armitage_018}. This implies that in Weyl fermions, the projection of spin over their momentum direction is preserved, a condition referred to as “spin-momentum locking”. Some remarkable properties related to the existence of Weyl nodes in the bulk band structure are the presence of Fermi arcs, the chiral anomaly, and the chiral magnetic effect~\cite{Burkov_018,Armitage_018}. Therefore, considerable effort has been devoted to the study of the electronic and thermoelectric transport properties of WSMs~\cite{Garcia_020,Han_020,Zhou_019,Yang_023,Markov_018,Bonilla_022,Bonilla_2024}, including the effects of different scattering mechanisms, such as electron–phonon, localized impurities and strain~\cite{Bonilla_2024}. 

The unique features of the band structure and its corresponding quasiparticle excitations in WSMs suggest the potential existence of diverse pairing states~\cite{Murakami_03,Meng_012,Cho-2012,Li_018,Sun_020}. Among them, the monopole superconductor (MSC), recently proposed as a pairing state between the two Fermi surfaces (FSs) enclosing the Weyl points in a doped WSM~\cite{Li_018,Sun_020}, involves the superposition of the Berry phases of the individual, single-particle states, thus determining the Berry phase of the emerging Cooper pair. Therefore, the resulting chirality, i.e. the vorticity of the pair, will be characterized by the combination of those of the individual Weyl nodes. As it was showed in \cite{Li_018}, for spherically symmetric Fermi surfaces, the pairing will be defined by monopole harmonic functions in contrast to the spherical harmonics in usual superconductivity. In order to experimentally detect those non-trivial vorticity effects, magneto-transport measurements can be applied to discriminate chiral versus non-chiral pairing states. This mechanism has been investigated by some of us~\cite{Munoz-2020} in the context of the competition between the monopole pairing, characterized by the monopole harmonic functions $\mathcal{Y}_{q,j,m}(\theta,\varphi)$, with conventional spherical harmonic states $Y_{j,m}(\theta,\varphi)$. In this previous work, within a mean-field BCS theory, we showed that the monopole  and a conventional spherical harmonic phase may coexist with one another~\cite{Munoz-2020}, while exclusion of one phase in favor of the other arises when the $\theta$-dependent form factors of the monopole harmonic $\mathcal{Y}_{|q|,j,|m|}(\theta,\varphi)$ and the spherical harmonic $Y_{j,m}(\theta,\varphi)$ are proportional to each other. We called this mechanism ``topological repulsion"~\cite{Munoz-2020}.

In this work, we expanded the analysis in Ref.~\cite{Munoz-2020} by introducing a generic microscopic model for the two-particle interaction potential~\cite{Cho-2012}, to self-consistently derive the angular dependence of the BCS order parameters involved in the inter-nodal and intra-nodal pairings leading to the MSC and conventional spherical harmonic (SHSC) superconducting phases, respectively. From this more detailed, microscopic perspective, we revisited the phase diagram predicted by a BCS mean-field theory~\cite{Munoz-2020}, thus determining the corresponding critical temperatures for the MSC and SHSC in terms of the microscopic interaction potential parameters and its underlying symmetries. We calculated explicitly the specific heat, and showed how the temperature behavior of this material dependent thermal property can provide a signature of the presence of such phases, as well as a direct experimental probe for their critical temperatures. We also investigated the competition between those phases, particularly concerning the topological repulsion mechanism suggested in our previous work~\cite{Munoz-2020}. In contrast with this previous work, where disorder effects were included as an asymmetry in the nodal chemical potentials, here we assume a uniform chemical potential. Finally, we provide some explicit examples for different angular dependence's of the pairing, as inherited by the underlying microscopic interaction potential symmetries.

\section{The Model}
\label{Sec:Model}
Let us start from the general form of the Hamiltonian representing a Weyl semimetal in the presence of particle-particle interactions
\begin{equation}        
    \hat{H}=\hat{H}_{0}+\hat{V}.
\end{equation}
Here, the non-interacting contribution is defined by
\begin{equation}
\hat{H}_{0}=\sum_{\mathbf{k},\sigma,\sigma'}\hat{c}^\dagger_{\sigma,\mathbf{k}}\left[\hat{H}_\mathbf{k}\right]_{\sigma\sigma'}\hat{c}_{\sigma',\mathbf{k}},
\label{eq_H0}
\end{equation}
for fermion operators $[ \hat{c}_{\sigma,\mathbf{k}},\hat{c}^\dagger_{\sigma',\mathbf{k}'}  ]_{+} =\delta_{\mathbf{k},\mathbf{k}'}\delta_{\sigma,\sigma'}$. The matrix in Eq.~\eqref{eq_H0} is defined, from a standard tight-binding model~\cite{Cho-2012} in Bloch-momentum space, by
\bea
    \hat{H}_\mathbf{k}&=&t(\hat{\sigma}^x\sin k_x+\hat{\sigma}^y\sin k_y)+t_z(\cos k_z-\cos Q)\hat{\sigma}^z\nn\\
    &&+m(2-\cos k_x-\cos k_y)\hat{\sigma}^z-\mu,
    \label{eq_Hk}
\eea
with $\hat{\sigma}^{(x,y,z)}$ the usual Pauli matrices. This model involves two Weyl points located at momenta $\mathbf{K}_\pm=(0,0,\pm Q)$, respectively.

In addition, we shall assume a short-range interaction, which includes a nearest-neighbors Coulomb repulsion $V_1>0$ and a contact, phonon-mediated effective attractive interaction $V_0<0$. When both terms combined we have
\begin{equation}
    \hat{V}=V_0\sum_i\hat{n}_i\hat{n}_i+\frac{1}{2}V_1\sum_{\langle i,j\rangle}\hat{n}_i\hat{n}_j.
    \label{eq_V}
\end{equation}
Here $\hat{n}_i=\sum_{\sigma}\hat{c}^\dagger_\sigma(\mathbf{R}_i)\hat{c}_\sigma(\mathbf{R}_i)$ is the occupation number operator at the site $\mathbf{R}_i$ on the lattice in the local Wannier basis. By the appropriate Fourier transformation $\hat{c}_\sigma(\mathbf{R}_i)=\frac{1}{\sqrt{N}}\sum e^{i\mathbf{k}\cdot\mathbf{R}_i}\hat{c}_\sigma(\mathbf{k})$, the interaction term is expressed in the Bloch-momentum basis by
\begin{equation}
\hat{V}=\sum_{\mathbf{k},\mathbf{p},\mathbf{q},\sigma,\tau}V_\mathbf{q}\hat{c}^\dagger_{\sigma,\mathbf{k}+\mathbf{q}}\hat{c}_{\sigma,\mathbf{k}}\hat{c}^\dagger_{\tau,\mathbf{p}-\mathbf{q}}\hat{c}_{\tau,\mathbf{p}},
\label{eq:original_momentum_interaction}
\end{equation}
with a matrix element given by
\begin{equation}
    V_\mathbf{q}=V_0+V_1(\cos q_x+\cos q_y+\cos q_z).
\label{eq:lattice_potential}
\end{equation}

In order to capture the topological features of the Weyl semimetal band structure, we define operators in the nodal Fermi surface representation, by expanding the momenta in the vicinity of each Weyl point as follows $\hat{\psi}_{\pm,\sigma}(\mathbf{k})=\hat{c}_\sigma(\mathbf{k}+\mathbf{K}_\pm)$ for $|\mathbf{k}|\ll|\mathbf{K}_\pm|$. By further assuming isotropy of the Fermi velocity around each of the Weyl points, such that $t=t_z\sin(Q)\equiv v_F$, we obtain the effective low-energy Hamiltonian
\bea
\hat{H}_{0}&=&\sum_{\mathbf{k},a=\pm}\hat{\psi}^\dagger_a(\mathbf{k})\big[v_F(k_x\hat{\sigma}^x+k_y\hat{\sigma}^y-a k_z\hat{\sigma}^z)-\mu\big]\hat{\psi}_a(\mathbf{k}),\nn\\
\eea
where $a=\pm$ represents the nodal topological charge (chirality).

For the interaction operator, we take a similar approach, starting from the Bloch-momentum representation (\ref{eq:original_momentum_interaction}) and choosing the pairing mechanism of our interest. For instance, for a BCS-like channel, the momentum of both particles will be opposite to each other in the center of mass frame. Specifying this condition, we obtain
\bea
 \hat{V}_{\text{BCS}}=\sum_{\mathbf{k},\mathbf{q},\sigma,\tau}V_{\mathbf{k}-\mathbf{q}}\hat{c}^\dagger_{\sigma,\mathbf{k}}\hat{c}^\dagger_{\tau,-\mathbf{k}}\hat{c}_{\tau,-\mathbf{q}}\hat{c}_{\sigma,\mathbf{q}}.
\eea
Then, we turn to the nodal Fermi surface representation by expanding the two momenta in terms of $\mathbf{K}_\pm$ as $\mathbf{k}=\mathbf{k'}+\mathbf{K}_a$ and $\mathbf{q}=\mathbf{q'}+\mathbf{K}_b$, with $a,b=\pm$ respectively. After this substitution, we obtain
\bea
\hat{V}_{\text{BCS}}=\sum_{\mathbf{k'},\mathbf{q'},\sigma,\tau,a,b}&&V_{\mathbf{k'}-\mathbf{q'}+\mathbf{K}_a-\mathbf{K}_b}\hat{\psi}^\dagger_{\sigma,a}(\mathbf{k'})\hat{\psi}^\dagger_{\tau,-a}(-\mathbf{k})\nn\\
    &&\times\hat{\psi}_{\tau,-b}(-\mathbf{q'})\hat{\psi}_{\sigma,b}(\mathbf{q'}),
\eea
and by assigning each possible value of the nodal indexes $a$ and $b$, we finally obtain
\begin{equation}  \hat{V}_{\text{BCS}}=\sum_{\mathbf{k},\mathbf{q},\sigma,\tau,a} V^{ab,cd}_{\mathbf{k},\mathbf{q}}\hat{\psi}^\dagger_{\sigma,a}(\mathbf{k})\hat{\psi}^\dagger_{\tau,b}(-\mathbf{k})\hat{\psi}_{\tau,c}(-\mathbf{q})\hat{\psi}_{\sigma,d}(\mathbf{q}),
\end{equation}
where we identify the coefficients
\begin{align*}
    V^{+-,-+}_{\mathbf{k},\mathbf{q}}&\,=V^{-+,+-+}_{\mathbf{k},\mathbf{q}}= V_{\mathbf{k}-\mathbf{q}},\\
    V^{+-,+-}_{\mathbf{k},\mathbf{q}}&\,=V_{\mathbf{k}-\mathbf{q}+2\mathbf{K}_+},\nn \\
    V^{-+,-+}_{\mathbf{k},\mathbf{q}}&\,=V_{\mathbf{k}-\mathbf{q}+2\mathbf{K}_-},
\end{align*}
corresponding to all the non-vanishing elements of the inter-node interaction. Now, to take care of the intra-node interactions, which possess a center of momentum at $2\mathbf{K}_\pm$, we choose the configuration $\mathbf{k}+\mathbf{p}=2\mathbf{K}_a$, such that the potential term, in analogy with the Fulde–Ferrell–Larkin–Ovchinnikov (FFLO) phase in conventional superconductors, reduces in this case to
\begin{equation}
\hat{V}_{\text{FFLO}}=\sum_{\mathbf{k},\mathbf{q},\sigma,\tau,a}V_{\mathbf{k}-\mathbf{q}}\hat{c}^\dagger_{\sigma,\mathbf{k}}\hat{c}^\dagger_{\tau,-\mathbf{k}+2\mathbf{K}_a}\hat{c}_{\tau,-\mathbf{q}+2\mathbf{K}_a}\hat{c}_{\sigma,\mathbf{q}}.
\end{equation}
Notice that in this second case, the nodal Fermi surface representation implies that both momenta are $\mathbf{k}=\mathbf{k'}+\mathbf{K}_a$ and $\mathbf{q}=\mathbf{q'}+\mathbf{K}_a$, near the same Weyl node $a$. With this substitution, this second channel turns out to be, 
\begin{equation}
\hat{V}_{\text{FFLO}}=\sum_{\mathbf{k},\mathbf{q},\sigma,\tau,a}V_{\mathbf{k}-\mathbf{q}}\hat{\psi}^\dagger_{\sigma,a}(\mathbf{k})\hat{\psi}^\dagger_{\tau,a}(-\mathbf{k})\hat{\psi}_{\tau,a}(-\mathbf{q})\hat{\psi}_{\sigma,a}(\mathbf{q}),
\end{equation}
where we identify the remaining non-zero potential coefficients as
\begin{equation}
    V^{++,++}_{\mathbf{k},\mathbf{q}}=V^{--,--}_{\mathbf{k},\mathbf{q}}=V_{\mathbf{k}-\mathbf{q}}.
\end{equation}
Finally, adding up both terms we obtain the total interaction potential
\bea
\hat{V}&=&\hat{V}_{\text{BCS}}+\hat{V}_{\text{FFLO}}\\
&=&\sum_{\mathbf{k},\mathbf{q},\mathbf{p}}V^{ab,cd}_\mathbf{q}\hat{\psi}^\dagger_{a,\sigma}(\mathbf{k+q})\hat{\psi}^\dagger_{b,\tau}(\mathbf{p-q})\hat{\psi}_{c,\tau}(\mathbf{p})\hat{\psi}_{d,\sigma}(\mathbf{k}),\nn
\eea
where now we include all the interaction coefficients obtained for both channels in a single matrix element $V^{ab,cd}_\mathbf{q}$. Explicitly, expanding up to second order in the momenta, we obtain
\begin{widetext}
\begin{align}
V^{\pm\pm,\pm\pm}_{\mathbf{k},\mathbf{q}}&\,=V^{\pm\mp,\mp\pm}_{\mathbf{k},\mathbf{q}}=V_0+V_1\bigg(3-\frac{(\mathbf{k}-\mathbf{q})^2}{2}\bigg)\nn\\
V^{\pm\mp,\pm\mp}_{\mathbf{k},\mathbf{q}}&=V_0+V_1\bigg\{2-\frac{|\mathbf{k}_\perp-\mathbf{q}_\perp|^2}{2}+\Big(1-\frac{(k_z-q_z)^2}{2}\Big)\cos2Q\mp(k_z-q_z)\sin 2Q\bigg\},
    \label{eq:Vcoeff}
\end{align}
\end{widetext}
with $\mathbf{k}_\perp=(k_x,k_y,0)$.  The proper treatment of these coefficients in the low-energy regime leads to the emergence of different possible superconducting phases of the model, according to its intra- or inter-node pairing structure. This analysis will be presented in Section~\ref{Sec:pairing_channels}, but first we shall establish the nature of the possible pairings. In a standard BCS-like mean-field approximation, we write the interaction as
\begin{equation}
    \hat{V}=\sum_{\mathbf{k},\sigma,\tau,a,b}\hat{\psi}^\dagger_{a,\sigma}(\mathbf{k})\Delta^{ab}_{\sigma\tau}(\mathbf{k})\hat{\psi}^\dagger_{b,\tau}(-\mathbf{k})+h.c.,
\end{equation}
where the pairing function is defined via the self-consistent relation
\begin{equation}
    \Delta^{ab}_{\sigma\tau}(\mathbf{k})=\sum_{\mathbf{q},c,d}V^{ab,cd}_{\mathbf{k},\mathbf{q}}\langle\hat{\psi}_{c,\tau}(-\mathbf{q})\hat{\psi}_{d,\sigma}(\mathbf{q})\rangle.
\end{equation}

Following the method presented in~\cite{Munoz-2020}, we introduce the Bogoliubov transformation $\hat{\alpha}^\dagger_a(\mathbf{k})=\sum_{\sigma}\xi_{a,\sigma}(\mathbf{k})\hat{\psi}^\dagger_{a,\sigma}(\mathbf{k})$, where the spinor $\xi_a(\mathbf{k})$ is the positive energy eigenvector of $\hat{H}_0$ at each Weyl node $a=\pm$. In spherical coordinates, we thus have (see Appendix~\ref{App:Bogoliubov} for details)
\begin{equation}
    \xi_+(\mathbf{k})=\begin{pmatrix}\sin(\theta_{k}/2)\\\cos(\theta_{k}/2)e^{i\phi_k}\end{pmatrix},\,\,\,\xi_-(\mathbf{k})=\begin{pmatrix}\cos(\theta_{k}/2)\\\sin(\theta_{k}/2)e^{i\phi_k}\end{pmatrix}.
\label{eq:bogoliubov_spinors}
\end{equation}
In this basis, the Hamiltonian is then reduced to
\begin{equation}
    \hat{H}=\sum_{\mathbf{k},a,b}\Big\{h_\mathbf{k}\hat{\alpha}^\dagger_a(\mathbf{k})\hat{\alpha}_a(\mathbf{k})+\Delta^{ab}(\mathbf{k})\hat{\alpha}^\dagger_a(\mathbf{k})\hat{\alpha}^\dagger_b(-\mathbf{k})+h.c.\Big\},
\end{equation}
where the single-particle contribution was diagonalized as $h_\mathbf{k}=v_F|\mathbf{k}|-\mu$, and the projected gap equation in the new basis becomes
\begin{equation}
    \Delta^{ab}(\mathbf{k})=\sum_{\mathbf{q},c,d}\Bar{V}^{ab,cd}_{\mathbf{k},\mathbf{q}}\langle\hat{\alpha}_c(-\mathbf{q})\hat{\alpha}_d(\mathbf{q})\rangle,
\label{eq:self_consistent_relation}
\end{equation}
with matrix elements $\Bar{V}^{ab,cd}_{\mathbf{k},\mathbf{q}}$ given by the corresponding transformation from the originals to this new basis.

Let us further define the Nambu spinor $\Psi^\dagger_\mathbf{k}=\big(\alpha^\dagger_+(\mathbf{k}),\hat{\alpha}_+(-\mathbf{k}),\hat{\alpha}^\dagger_-(\mathbf{k}),\hat{\alpha}_-(-\mathbf{k})\big)$, in order to arrange the Hamiltonian in the Bogoliubov-de Gennes notation as a bilinear form
\begin{equation}
\hat{H}=\sum_\mathbf{k}\Psi^\dagger_\mathbf{k}\hat{H}_{\text{BdG}}(\mathbf{k})\Psi_\mathbf{k},
\end{equation}
where we defined the matrix
\begin{equation}
    \hat{H}_{\text{BdG}}(\mathbf{k})=
    \begin{pmatrix}
    h_\mathbf{k} & \Delta_0(\mathbf{k}) & 0 & \Delta_1(\mathbf{k})
    \\
    \Delta_0^*(\mathbf{k}) & -h_\mathbf{k} & \Delta_1^*(\mathbf{k}) & 0
    \\
    0 & \Delta_1(\mathbf{k}) & h_\mathbf{k} & \Delta_0(\mathbf{k})
    \\
    \Delta_1^*(\mathbf{k}) & 0 & \Delta_0^*(\mathbf{k}) & -h_\mathbf{k}
    \end{pmatrix}.
\end{equation}

For subsequent computations, it is convenient to express this matrix in a $\text{SU}(2)\otimes\text{SU}(2)$ basis composed by $\{\hat{\tau}_\alpha\otimes\hat{\eta}_\beta\}$, where $\hat{\tau}_\alpha$ represents the two Weyl nodes subspace, while $\hat{\eta}_\beta$ represent the particle-hole degree of freedom. Here, $\alpha,\beta=1,2,3$ represent the Pauli matrix indexes, while $\alpha,\beta=0$ indexes the two-dimensional representation of the identity. With this notation, we obtain
\bea
    &&\hat{H}_{\text{BdG}}(\mathbf{k})= h_\mathbf{k}\hat{\tau}_0\otimes\hat{\eta}_3+\text{Re}\Delta_0(\mathbf{k})\hat{\tau}_0\otimes\hat{\eta}_1\\
    &&-\text{Im}\Delta_0(\mathbf{k})\hat{\tau}_0\otimes\hat{\eta}_2+\text{Re}\Delta_1(\mathbf{k})\hat{\tau}_1\otimes\hat{\eta}_1-\text{Im}\Delta_1(\mathbf{k})\hat{\tau}_1\otimes\hat{\eta}_2.\nn
\eea
\\
\section{Gap Equations}
\label{Sec:Gap}
Let us now consider the functional integral representation of the partition function at finite temperature $T = \beta^{-1}$,
\begin{equation}
Z=\int\mathcal{D}\Psi^\dagger_\mathbf{k}\mathcal{D}\Psi_\mathbf{k}e^{-S[\Psi^\dagger_\mathbf{k},\Psi_\mathbf{k}]},
\end{equation}
where the action in the Euclidean time $\tau\in[0,\beta]$ is given by
\begin{equation}
S\big[\Psi^\dagger_\mathbf{k},\Psi_\mathbf{k}\big]=\int_0^\beta d\tau \Psi^\dagger_\mathbf{k}(\tau)\left(\partial_\tau+\hat{H}_\text{BdG}(\mathbf{k})\right)\Psi_\mathbf{k}(\tau),
\end{equation}
and the integration is performed over the Grassmann fields $\Psi_\mathbf{k},\Psi_\mathbf{k}^\dagger$ with anti-perodic boundary conditions $\Psi_\mathbf{k}(\beta)=-\Psi_\mathbf{k}(0)$. Therefore, the Green's function of the system is obtained as the resolvent
\begin{equation}
\left(\partial_\tau+\hat{H}_\text{BdG}(\mathbf{k})\right)\hat{\mathcal{G}}_\mathbf{k}(\tau)=\delta(\tau).
\end{equation}
It is more convenient to solve this equation in the Matsubara frequency space, where $\partial_\tau\to-i\omega_n$ (for $\omega_n = (2n + 1)\pi/\beta,\,\,n\in\mathbb{Z}$), so we can write
\bea
    &&\hat{\mathcal{G}}_\mathbf{k}(\omega_n)=\big[-i\omega_n\hat{\tau}_0\otimes\hat{\eta}_0+\hat{H}_{\text{BdG}}(\mathbf{k})\big]^{-1}.
\label{eq:green_f_to_invert}
\eea

Therefore, the Green's function corresponds to the block-matrix form
\begin{equation}
    \hat{\mathcal{G}}_\mathbf{k}(\omega_n)=\begin{pmatrix}
    \hat{\mathcal{G}}^{++}_\mathbf{k}(\omega_n) & \hat{\mathcal{G}}^{+-}_\mathbf{k}(\omega_n) \\
    \hat{\mathcal{G}}^{-+}_\mathbf{k}(\omega_n) & \hat{\mathcal{G}}^{--}_\mathbf{k}(\omega_n)
    \end{pmatrix}.
\end{equation}
The explicit expressions for each of the four matrix components are
\begin{widetext}
\bea
\hat{\mathcal{G}}^{\pm\pm}_\mathbf{k}(\omega_n)&=&\frac{A_\mathbf{k}i\omega_n\hat{\eta}_0+\big(A_\mathbf{k}\text{Re}\Delta_0(\mathbf{k})-2B_\mathbf{k}\text{Re}\Delta_1(\mathbf{k})\big)\hat{\eta}_1+\big(2B_\mathbf{k}\text{Im}\Delta_1(\mathbf{k})-A_\mathbf{k}\text{Im}\Delta_0(\mathbf{k})\big)\hat{\eta}_2+A_\mathbf{k}h_\mathbf{k}\hat{\eta}_3}{A_\mathbf{k}^2-4B_\mathbf{k}^2}
\nn\\
\hat{\mathcal{G}}^{\pm\mp}_\mathbf{k}(\omega_n)&=&\frac{-2B_\mathbf{k}i\omega_n\hat{\eta}_0+\big(A_\mathbf{k}\text{Re}\Delta_1(\mathbf{k})-2B_\mathbf{k}\text{Re}\Delta_0(\mathbf{k})\big)\hat{\eta}_1+\big(2B_\mathbf{k}\text{Im}\Delta_0(\mathbf{k})-A_\mathbf{k}\text{Im}\Delta_1(\mathbf{k})\big)\hat{\eta}_2-2B_\mathbf{k}h_\mathbf{k}\hat{\eta}_3}{A_\mathbf{k}^2-4B_\mathbf{k}^2},
\eea
\end{widetext}
where we defined the variables
\bea
A_\mathbf{k}&=&\omega_n^2+h_\mathbf{k}^2+|\Delta_0(\mathbf{k})|^2+|\Delta_1(\mathbf{k})|^2\equiv\omega_n^2+a_\mathbf{k}^2\nn\\ B_\mathbf{k}&=&\text{Re}\Delta_0(\mathbf{k})\text{Re}\Delta_1(\mathbf{k})+\text{Im}\Delta_0(\mathbf{k})\text{Im}\Delta_1(\mathbf{k}).
\eea
With this result, we can now complete the self-consistent equation for both pairing gaps as follows
\begin{equation}
    \Delta_0(\mathbf{k})=\frac{1}{\beta}\sum_{\omega_n,\mathbf{q}}\Bar{V}^{++,++}_{\mathbf{k},\mathbf{q}}\langle\hat{\alpha}_+(-\mathbf{q},\omega_n)\hat{\alpha}_+(\mathbf{q},\omega_n)\rangle
\end{equation}
and
\bea
    \Delta_1(\mathbf{k})&=&\frac{1}{\beta}\sum_{\omega_n,\mathbf{q}}\big(\Bar{V}^{+-,+-}_{\mathbf{k},\mathbf{q}}+\Bar{V}^{+-,-+}_{\mathbf{k},\mathbf{q}}\big)\nonumber\\
    &&\times\langle\hat{\alpha}_+(-\mathbf{q},\omega_n)\hat{\alpha}_-(\mathbf{q},\omega_n)\rangle.
\eea
Here, we identify each of the correlators with a matrix element of the corresponding nodal Green function, as follows
\bea
    \langle\hat{\alpha}_+(-\mathbf{q},\omega_n)\hat{\alpha}_+(\mathbf{q},\omega_n)\rangle &=&[\mathcal{G}^{++}_\mathbf{k}(\omega_n)]_{21}\\
&=&\frac{A_\mathbf{k}\Delta_0(\mathbf{k})-2B_\mathbf{k}\Delta_1(\mathbf{k})}{A^2_\mathbf{k}-4B^2_\mathbf{k}},\nn
\eea
and
\bea
    \langle\hat{\alpha}_+(-\mathbf{q},\omega_n)\hat{\alpha}_-(\mathbf{q},\omega_n)\rangle &=& [\mathcal{G}^{+-}_\mathbf{k}(\omega_n)]_{21}\\
    &=&\frac{A_\mathbf{k}\Delta_1(\mathbf{k})-2B_\mathbf{k}\Delta_0(\mathbf{k})}{A^2_\mathbf{k}-4B^2_\mathbf{k}}.\nn
\eea

After computing the Matsubara sums by standard methods, we obtain the coupled system of BCS gap equations
\begin{widetext}
\bea
\Delta_0(\mathbf{k})&=&\sum_{\mathbf{q}}\Bar{V}^{++,++}_{\mathbf{k},\mathbf{q}}\Bigg\{\frac{\big(\Delta_0(\mathbf{q})+\Delta_1(\mathbf{q})\big)}{4E_+}\tanh(\beta E_+/2)+\frac{\big(\Delta_0(\mathbf{q})-\Delta_1(\mathbf{q})\big)}{4E_-}\tanh(\beta E_-/2)\Bigg\}\nn\\  \Delta_1(\mathbf{k})&=&\sum_{\mathbf{q}}\big(\Bar{V}^{+-,+-}_{\mathbf{k},\mathbf{q}}+\Bar{V}^{+-,-+}_{\mathbf{k},\mathbf{q}}\big)\Bigg\{\frac{\big(\Delta_1(\mathbf{q})+\Delta_0(\mathbf{q})\big)}{4E_+}\tanh(\beta E_+/2)+\frac{\big(\Delta_1(\mathbf{q})-\Delta_0(\mathbf{q})\big)}{4E_-}\tanh(\beta E_-/2)\Bigg\},
\eea
\end{widetext}
where we defined the excitation energies $E^2_\pm=a_\mathbf{k}^2\pm2B_\mathbf{k}$. Notice that these equations involve the presence of two order parameters, i.e. $\Delta_0$ and $\Delta_1$. Moreover, if we take the limit $\Delta_1\to0$, the energies $E_\pm\to\sqrt{h_\mathbf{k}^2+|\Delta_0(\mathbf{k})|^2}$, and we recover the usual BCS equation. 

We shall consider the continuum limit $\sum_\mathbf{k}\to\frac{\mathcal{V}}{(2\pi)^3}\int d^3k$, and hence it is convenient to introduce the density of states in momentum integrals such as
\bea
    \int\frac{d^3q}{(2\pi)^3}f(E_q,\Omega_\mathbf{q})&=&\int d\Omega_\mathbf{q}\int^{\omega_D}_{-\omega_D} dE\rho(E)f(E,\Omega_\mathbf{q})\nn\\
    &\approx &\rho(\mu)\int d\Omega_\mathbf{q}\int^{\omega_D}_{-\omega_D}f(E,\Omega_\mathbf{q}),
\label{eq:dos}
\eea
where $f(E_q,\Omega_q)$ is any function that only depends on $|\mathbf{q}|$ through $E_q$, $\omega_D$ is the Debye frequency, and the density of states is defined by
\begin{equation}
    \rho(E)\equiv \int\frac{d^3q}{(2\pi)^3}\delta(E-v_F|\mathbf{q}|+\mu)=\frac{1}{4\pi^2}\frac{(E+\mu)^2}{v_F^3}.
\end{equation}
In the last step of Eq.~\eqref{eq:dos}, we consider that the fermionic states are very dense in the neighborhood of the Fermi level $\mu$.

We further expand the interaction potential matrix elements according to their angular dependence, as follows
\bea
&&\Bar{V}^{++,++}_{\mathbf{k},\mathbf{q}}\equiv \Bar{V}^{\text{intra}}_{lm} Y_{l,m}(\theta_\mathbf{k},\phi_\mathbf{k})Y^*_{l,m}(\theta_\mathbf{q},\phi_\mathbf{q})\nn\\
&&\Bar{V}^{+-,+-}_{\mathbf{k},\mathbf{q}}+\Bar{V}^{+-,-+}_{\mathbf{k},\mathbf{q}} \equiv \Bar{V}^{\text{inter}}_{q,l,m} \mathcal{Y}_{q,l,m}(\theta_\mathbf{k},\phi_\mathbf{k})\mathcal{Y}^*_{q,l,m}(\theta_\mathbf{q},\phi_\mathbf{q}),\nn\\
\label{eq:VHS} 
\eea
where the explicit coefficients  $\Bar{V}^{\text{intra}}_{lm}$ and $\Bar{V}^{\text{inter}}_{q,l,m}$ in Eq.~\eqref{eq:VHS} are determined from the expansion of Eqs.~\eqref{eq:Vcoeff} in spherical $Y_{l,m}(\theta,\phi)$ and monopole $\mathcal{Y}_{q,l,m}(\theta,\phi)$ harmonics, respectively. Some explicit examples will be presented in Sec.~\ref{Sec:pairing_channels}.

Finally, since the self-consistent Eq.~\eqref{eq:self_consistent_relation} implies that the angular dependence of the pairing gap functions must match the corresponding interaction potential, we expand each gap pairing function as follows
\bea
&&\Delta_0(\mathbf{q})=\Delta^{l,m}_0 Y_{l,m}(\theta_\mathbf{q},\phi_\mathbf{q})\equiv \Delta^{l,m}_0 f_{l,m}(\theta_\mathbf{q})e^{im\phi_\mathbf{q}},\nn\\
&&\Delta_1(\mathbf{q})=\Delta^{q,l,m}_1\mathcal{Y}_{q,l,m}(\theta_\mathbf{q},\phi_\mathbf{q})\equiv \Delta^{q,l,m}_1 g_{q,l,m}(\theta_\mathbf{q})e^{i(m+q)\phi_\mathbf{q}}.\nn\\
\label{pairing_gap_expansion}
\eea

Implementing all these structures in the BCS equations, we obtain the coupled integral equation system
\begin{widetext}
\begin{equation}
    \frac{1}{\lambda_0}=\int d\Omega_\mathbf{q}\int^{\omega_D}_{-\omega_D}dEY^*_{lm}(\Omega_\mathbf{q})\Bigg\{\frac{Y_{l'm'}(\Omega_\mathbf{q})}{2}\big(\mathcal{T}_\beta(E_+)+\mathcal{T}_\beta(E_-)\big)+u\frac{\mathcal{Y}_{q'l'm'}(\Omega_\mathbf{q})}{2}\big(\mathcal{T}_\beta(E_+)-\mathcal{T}_\beta(E_-)\big)\Bigg\}
\label{gap_eq_intra}
\end{equation}
and
\begin{align}
    \frac{1}{\lambda_1}=\int d\Omega_\mathbf{q}\int^{\omega_D}_{-\omega_D}dE\mathcal{Y}^*_{qlm}(\Omega_\mathbf{q})\Bigg\{&\frac{\mathcal{Y}_{q'l'm'}(\Omega_\mathbf{q})}{2}\big(\mathcal{T}_\beta(E_+)+\mathcal{T}_\beta(E_-)\big)+\frac{Y_{l',m'}(\Omega_\mathbf{q})}{2u}\big(\mathcal{T}_\beta(E_+)-\mathcal{T}_\beta(E_-)\big)\Bigg\},
\label{gap_eq_inter}
\end{align}
\end{widetext}
where we defined the function $\mathcal{T}_\beta(E)=\frac{\tanh(\beta E/2)}{2E}$, the effective coupling constants $\lambda_0=\rho(\mu)V^{\text{intra}}_{lm}$ and $\lambda_1=\rho(\mu)V^{\text{inter}}_{qlm}$, and the parameter $u=\Delta_1^{q'l'm'}/\Delta_0^{lm}$. Notice that the angular dependence is still present in the energy dispersion relations since
\begin{widetext}
\begin{equation}
    E_\pm=\sqrt{E^2+\left(\Delta_0^{l,m}\right)^2f^2_{l,m}(\theta_\mathbf{q})+\left(\Delta_1^{q',l',m'}\right)^2g^2_{q',l',m'}(\theta_\mathbf{q})\pm2B_\mathbf{q}},
    \label{eq:Epm}
\end{equation}
and
\begin{align}
    B_\mathbf{q}&=\Delta_0^{lm}\Delta_1^{q'l'm'}f_{lm}(\theta_\mathbf{q})g_{q'l'm'}(\theta_\mathbf{q})\big[\cos(m\phi_\mathbf{q})\cos([m'+q']\phi_\mathbf{q})+\sin(m\phi_\mathbf{q})\sin([m'+q']\phi_\mathbf{q})\big]\nonumber\\
    &=\Delta_0^{lm}\Delta_1^{q'l'm'}f_{lm}(\theta_\mathbf{q})g_{q'l'm'}(\theta_\mathbf{q})\cos\big([m-m'-q']\phi_\mathbf{q}\big).
\end{align}
\end{widetext}

\subsection{Critical Behavior}
\label{Sec:Critical}
Along this section, we shall analyze the possible solutions to the coupled system of BCS equations Eq.~\eqref{gap_eq_inter} and Eq.~\eqref{gap_eq_intra}, that determine the possible existence of the superconducting phases in the WSM. Moreover, she shall determine the magnitude of the critical temperatures associated to each of those SC phases. For notational convenience, along this section we will write the gap equations as
\begin{equation}
    \frac{\Delta_\eta}{\lambda_\eta}=A_\eta \Delta_\eta +B_\eta\Delta_{\bar\eta},
\label{eq:gap_eq_critical_analysis}
\end{equation}
for $\eta=0,1$ and $\bar\eta=1-\eta$. Here we also defined the integrals
\begin{widetext}
\begin{align}
    A_\eta&\equiv A_\eta(\Delta_\eta,\Delta_{\bar\eta})=\int\frac{d\Omega}{4\pi}f^2_\eta(\theta)\int_0^{\omega_D}\frac{dE}{2}\bigg\{\frac{\tanh(\beta E_+/2)}{E_+}+\frac{\tanh(\beta E_-/2)}{E_-}\bigg\}, \nn\\
    B_\eta&\equiv B_\eta(\Delta_\eta,\Delta_{\bar\eta})=\int\frac{d\Omega}{4\pi}f^2_\eta(\theta)e^{\pm is\phi}\int_0^{\omega_D}\frac{dE}{2}\bigg\{\frac{\tanh(\beta E_+/2)}{E_+}-\frac{\tanh(\beta E_-/2)}{E_-}\bigg\}.
\end{align}
\end{widetext}
Let us start by assuming the existence of two critical temperatures, corresponding to the vanishing point of each of the gap functions, i.e. $\Delta_\eta(T)=0$ for $T>T_c^\eta$. Without loss of generality, we denote by $\eta$ the higher critical temperature, so that the condition $T_c^\eta>T_c^{\bar\eta}$ applies.

First, we take the limit $T\to T_c^{\eta+}$ (the higher critical temperature), in which case $B_\eta=B_\eta(0,0)=0$ and the energy dispersion in Eq.~\eqref{eq:Epm} reduces to $E_{\pm}\to E$. Therefore, in this limit we obtain
\begin{equation}
    \frac{1}{\lambda_\eta}=\int\frac{d\Omega}{4\pi}f^2_\eta(\theta)\int_0^{\omega_D}dE\frac{\tanh(\beta E/2)}{E}\approx \frac{1}{4\pi}\ln\bigg[\frac{2\omega_De^{\gamma}}{\pi T_c^\eta}\bigg],
\end{equation}
where $\gamma$ is the Euler-Mascheroni constant and in the last step we applied the weak coupling approximation $\omega_D\gg\Delta_\eta$. Then, we solve for the higher critical temperature to obtain
\begin{equation}
    T_c^\eta=\bigg(\frac{\omega_D e^\gamma}{2\pi^2}\bigg) e^{-4\pi/\lambda_\eta}.
    \label{eq:Tc}
\end{equation}
Let us now approach the same critical temperature from below, i.e. the limit $T\to T_c^{\eta-}$, by expanding the $A_\eta(\Delta_\eta,0)$ in terms of $\Delta_\eta\ll1$ as
\begin{align}
\mathcal{T}_\beta(\sqrt{E^2+\Delta_\eta^2f_\eta^2(\theta)})\approx \mathcal{T}_\beta(E)+\psi(\beta,\beta E/2)\frac{\Delta_\eta^2f_\eta^2(\theta)}{2},
\end{align}
where we defined the auxiliary function
\begin{equation}
    \psi(\beta,x)=\frac{\beta^3}{8}\bigg(\frac{1}{x^2\cosh^2(x)}-\frac{\tanh(x)}{x^3}\bigg).
    \label{eq_psi}
\end{equation}
Then, by this substitution and after a careful evaluation of the corresponding integrals, we obtain
\begin{equation}
    \frac{1}{\lambda_\eta}=\frac{1}{4\pi}\ln\bigg[\frac{2\omega_D e^\gamma}{\pi T}\bigg]-\frac{\alpha}{4\pi T^2}\Lambda_\eta^4 \Delta_\eta^2,
\end{equation}
where we defined the constant $\alpha=7\zeta(3)/8\pi^2$, with $\zeta(s)$ the Riemann's Zeta function, as well as the angular integral
\begin{equation}
    \Lambda_{\eta,\bar\eta}^{n,m}=\int_0^\pi d\theta\sin\theta |f_\eta(\theta)|^n|f_{\bar\eta}(\theta)|^m.
    \label{eq_Lambda}
\end{equation}
Here and from now on, when any of the upper-index values equals zero, we shall omit the corresponding sub-index to alleviate notation.

Finally, comparing with the expression for the critical temperature in Eq.~\eqref{eq:Tc}, we can write
\begin{equation}
\Delta^2_\eta\equiv\Delta^2_\eta(T)=\frac{T^2}{\alpha\Lambda_{\eta}^4}\ln(T_c^\eta/T)\sim |T-T_c^\eta|,
    \label{eq_Deltac_eta}
\end{equation}
from where we obtain a critical exponent $\beta=1/2$. These results are not surprising, since in this limit we just recover the usual behavior of BCS theory. 

Now, at the lower critical temperature $T_c^{\bar\eta}$, we have $\Delta_\eta\neq0$, and in order to obtain a non-trivial equation from Eq.~\eqref{eq:gap_eq_critical_analysis} by taking $\Delta_{\bar\eta}\to0$ we need to take care of the limit on the second term,
\begin{widetext}
\begin{align}
    \lim_{\Delta_{\bar\eta}\to0}\Delta_\eta\frac{B_{\bar\eta}}{\Delta_{\bar\eta}}&\,=\int\frac{d\Omega}{4\pi}\int_0^{\omega_D}dE e^{\pm i s\phi}f_\eta(\theta)f_{\bar\eta}(\theta)\lim_{\Delta_{\bar\eta}\to0}\frac{\mathcal{T}_\beta(E_+)-\mathcal{T}_\beta(E_-)}{\Delta_{\bar\eta}}\Delta_\eta  \nn\\
    &\,=\Delta_\eta^2(T_c^{\bar\eta})\int\frac{d\Omega}{4\pi}\int_0^{\omega_D}dE e^{\pm i s\phi}f^2_\eta(\theta)f^2_{\bar\eta}(\theta)\cos(s\phi)\,\psi\bigg(\beta_c^{\bar\eta},\beta_c^{\bar\eta}\frac{\sqrt{E^2+\Delta_\eta^2f_\eta^2(\theta)}}{2}\bigg),
    \label{eq_Blimit}
\end{align}
\end{widetext}
where $\psi(\beta,x)$ was defined in Eq.~\eqref{eq_psi}. Notice that the overall dependence on the azimuthal angle is integrated out to obtain
\begin{equation}
    \int_0^{2\pi}e^{\pm is\phi}\cos(s\phi)=\frac{1}{2}(\delta_{s,0}+1).
    \label{eq:azhimut}
\end{equation}
Now, taking the limit $T\to T_c^{\bar\eta+}$ in the first term in Eq.~\eqref{eq:gap_eq_critical_analysis}, we obtain
\begin{equation}
    A_\eta(\Delta_\eta,0)\approx\frac{1}{4\pi}\ln\bigg[\frac{2\omega_De^\gamma}{\pi T_c^{\bar\eta}}\bigg]-\frac{\alpha}{4\pi (T_c^{\bar\eta})^2}\Lambda_{\eta,\bar\eta}^{2,2}\Delta_\eta^2(T_c^{\bar\eta}).
\end{equation}
We shall further assume that both critical temperatures are close enough to each other in order to use the critical behavior of $\Delta_\eta(T)$ in the vicinity of $T_c^\eta$ as defined in Eq.~\eqref{eq_Deltac_eta}. This assumption is based on the fact that the coexistence of phases is expected to occur for couplings of similar values. With this assumption, we can expand the first term in Eq.~\eqref{eq_Blimit} up to second order in $\Delta_\eta$, to obtain
\begin{align}
    \frac{4\pi}{\lambda_{\bar\eta}}=\ln\bigg[\frac{2\omega_De^\gamma}{\pi T_c^{\bar\eta}}\bigg]-\frac{\alpha}{ (T_c^{\bar\eta})^2}(2+\delta_{s,0})\Lambda_{\eta,\bar\eta}^{2,2}\Delta_\eta^2(T_c^{\bar\eta}).
\end{align}
Thereby, replacing the argument in the logarithm in terms of $T_c^\eta$ in Eq.~\eqref{eq:Tc}, and using again the critical behavior of $\Delta_\eta$ from Eq.~\eqref{eq_Deltac_eta}, we obtain the expression
\bea
    4\pi\bigg(\frac{1}{\lambda_{\bar\eta}}-\frac{1}{\lambda_\eta}\bigg)=\nu_\eta \ln\bigg[\frac{T^\eta_c}{T_c^{\bar\eta}}\bigg],
     \label{eq:comp}
\eea
where we defined the parameters
\be
\nu_{\eta} = 1-(2+\delta_{s,0})\frac{\Lambda_{\eta,\bar\eta}^{2,2}}{\Lambda_\eta^4}.
\label{eq:nudef}
\ee
From Eq.~\eqref{eq:comp}, after elementary algebra, we conclude that the relation among both critical temperatures as a function of the corresponding couplings is given by
\begin{equation}
    T_c^{\bar\eta}=T_c^\eta e^{-\frac{4\pi}{\nu_\eta} \big( \frac{1}{\lambda_{\bar\eta}}-\frac{1}{\lambda_\eta}\big)}. 
\end{equation}
Finally, in order to obtain the critical behavior of both order parameters in the neighborhood of the second critical temperature, we expand every term of the gap equations up to second order in the pairings, as follows
\begin{align}
    A_\eta&\,\approx\frac{1}{4\pi}\ln\bigg[\frac{2\omega_De^\gamma}{\pi T}\bigg]-\frac{\alpha}{4\pi T^2}\big(\Lambda_\eta^4\Delta_\eta^2+\Lambda_{\eta,\bar\eta}^{2,2}\Delta_{\bar\eta}^2\big),\nn\\
    A_{\bar\eta}&\,\approx\frac{1}{4\pi}\ln\bigg[\frac{2\omega_De^\gamma}{\pi T}\bigg]-\frac{\alpha}{4\pi T^2}\big(\Lambda_\eta^4\Delta_{\bar\eta}^2+\Lambda_{\eta,\bar\eta}^{2,2}\Delta_\eta^2\big),\nn\\
    B_\eta&\,=B_{\bar\eta}\approx-\frac{\alpha(1+\delta_{s,0})}{T^2}\Lambda_{\eta,\bar\eta}^{2,2}\Delta_\eta\Delta_{\bar\eta}.
\end{align}
Then, replacing the coupling constants in terms of the critical temperatures, we finally solve the system of equations in favour of the pairing gaps, leading to the explicit expressions
\bea
    \Delta_\eta^2(T) &=&\frac{(T_c^{\bar\eta})^2}{\alpha\Lambda_\eta^4}\ln(T_c^\eta/T_c^{\bar\eta})\nn\\
    &+&\frac{T^2}{\alpha\Lambda^4_\eta}\bigg(\frac{\delta_\eta^2-(1-\nu_\eta)}{\delta^2_\eta-(1-\nu_\eta)^2}\bigg)\ln(T_c^{\bar\eta}/T),
    \label{eq:critical_gap_1}
\eea
and
\bea
    \Delta_{\bar\eta}^2(T)&\,=\frac{T^2}{\alpha\Lambda_{\bar\eta}^4}\bigg(\frac{\nu_\eta}{\delta^2_\eta-(1-\nu_\eta)^2}\bigg)\ln(T_c^{\bar\eta}/T),\label{eq:critical_gap_2}
\eea
where we defined the parameters
\be
\delta_\eta=\sqrt{\Lambda^4_{\bar\eta}/\Lambda^4_\eta}
\label{eq:deltadef}
\ee
in terms of the angular integrals in Eq.~\eqref{eq_Lambda}.

Notice that the first term in Eq.~\eqref{eq:critical_gap_1} is just the expression that we previously obtained in the neighborhood of $T_c^\eta$ in Eq.~\eqref{eq_Deltac_eta}, evaluated at $T=T_c^{\bar\eta}$, while the second term produces an additional contribution to its critical behavior. From this explicit result, we remark that the sign of the factor in parenthesis will determine whether the first pairing gap would increase or decrease along with the emergence of the second one. In addition, if we now inspect Eq.~\eqref{eq:critical_gap_2}, we find that a necessary condition to produce the second phase transition is
\begin{equation}
    \frac{\nu_\eta}{\delta_\eta^2-(1-\nu_\eta)^2}>0.
\end{equation}

We can analyze the combination of signs that determines the behavior of $\Delta_\eta$ in terms of $\nu_{\bar\eta}$, by combining Eqs.~\eqref{eq:nudef},~\eqref{eq:deltadef} to obtain the identity
\begin{align*}
    \delta_{\eta}^2-\left(1 -\nu_{\eta}\right) = \delta_{\eta}^2\,\nu_{\bar{\eta}},
\end{align*}
which implies the condition
\begin{equation}
    \rm{sgn}[\delta_{\eta}^2-(1-\nu_{\eta})]=\rm{sgn}[\nu_{\bar{\eta}}].
\end{equation}

All the conditions that we find necessary to produce the coexistence of both SC phases are summarized in Table~\ref{table_1}. As a final remark, we notice that we find a critical exponent $\beta=1/2$ for both pairing gaps, near each critical point, in agreement with the BCS-like theory that we developed here.

\begin{table}
\begin{tabular}{|c|c|c|c|c|}
     \hline
     \multicolumn{5}{|c|}{Conditions for Phase Coexistence with $T_c^\eta>T_c^{\bar \eta}$}\\\hline
     Coupling & Condition 1& Condition 2& Condition 3& Type  \\\hline
     \multirow{2}{2cm}{$\lambda_\eta>\lambda_{\bar \eta}$}& \multirow{2}{1cm}{$\nu_\eta>0$} & \multirow{2}{2cm}{$|1-\nu_\eta|<\delta_\eta$} & $\nu_{\bar \eta}>0$ & $\mathcal{I}$  \\\cline{4-5}
     & & & $\nu_{\bar \eta}<0$ & $\mathcal{D}$\\\hline
     \multirow{2}{2cm}{$\lambda_\eta<\lambda_{\bar \eta}$}& \multirow{2}{1cm}{$\nu_\eta<0$} & \multirow{2}{2cm}{$|1-\nu_\eta|>\delta_\eta$} & $\nu_{\bar \eta}>0$ & $\mathcal{D}$  \\\cline{4-5}
     & & & $\nu_{\bar \eta}<0$ & $\mathcal{I}$\\\hline
\end{tabular}
\caption{Summary of conditions required to generate the coexistence of superconducting phases. The last row denotes the increasing ($\mathcal{I}$) or decreasing ($\mathcal{D}$) behavior of the first gap at the second phase transition.}
\label{table_1}
\end{table}

\subsection{Specific Heat}
A thermal property that may provide a signature for the experimental characterization of the phase transitions discussed in our theory is the specific heat. Since the internal energy of the system is given by ($\beta = 1/T$)
\begin{equation}
    U=\langle\hat{H}\rangle=-\frac{\partial}{\partial\beta}\ln Z+\frac{\mu}{\beta}\frac{\partial}{\partial\mu}\ln Z,
\end{equation}
the specific heat is obtained as
\bea
    C_v=\frac{\partial}{\partial T}U=\left(\beta^2\frac{\partial^2}{\partial\beta^2}+\mu\frac{\partial}{\partial\mu}-\mu\beta\frac{\partial^2}{\partial\beta\partial\mu}\right)\ln Z.
\eea

\begin{figure}[h]
    \includegraphics[width=0.4\textwidth]{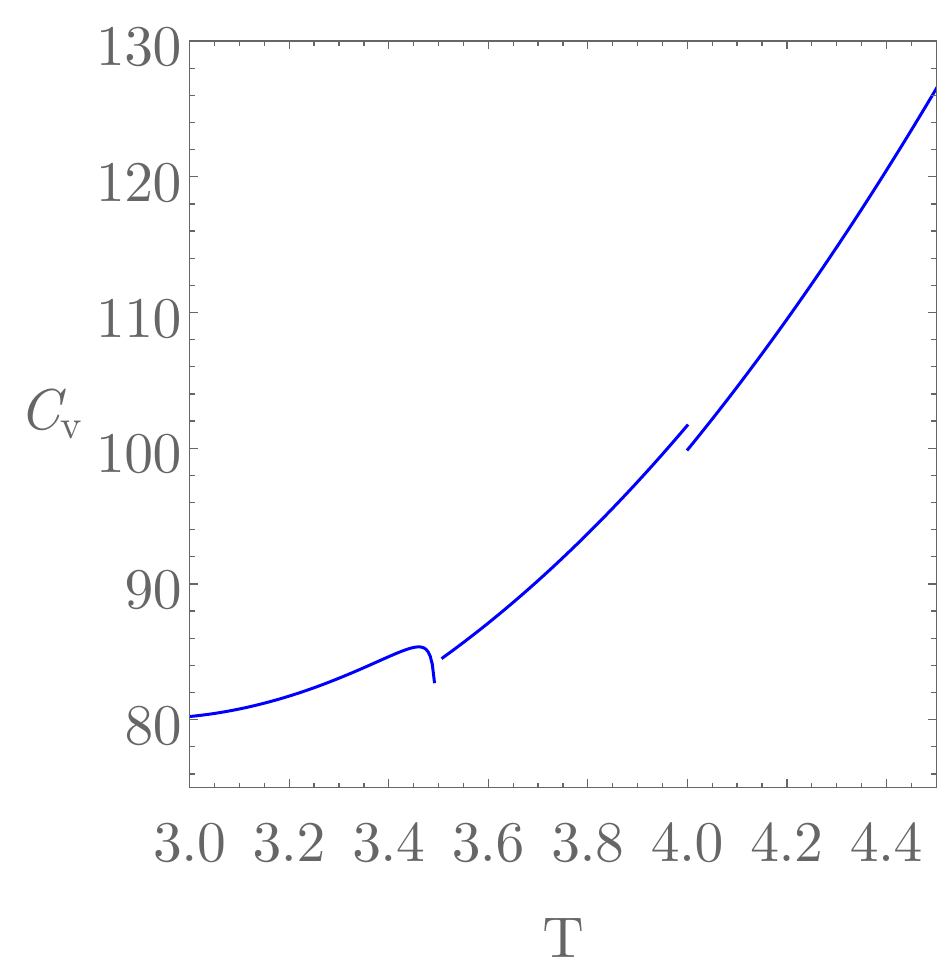}
    \caption{Total specific heat $C_v^{SC} + C_v^{norm}$ (in units of the Boltzmann constant $k_B$), from Eq.~\eqref{eq:cvSC} and Eq.~\eqref{eq:cvNorm}, respectively, as a function of temperature, at zero chemical potential $\mu=0$. The temperature is displayed in units of $\hbar v_F Q/k_B$.}
    \label{fig:specific_heat}
\end{figure}

The Grand Canonical partition function is obtained, upon functional integration over the Fermion fields, in terms of the Matsubara Green's function Eq.~\eqref{eq:green_f_to_invert}
\bea
    \ln Z &=&\ln\left(\text{det}[\hat{\mathcal{G}}^{-1}_\mathbf{k}(\omega_n)]\right)\nn\\ 
    &=&\sum_{\mathbf{k},\omega_n}\sum_{\zeta=\pm}\ln\left[\left( \ii \omega_n \right)^2 - E_{\zeta}^2\right]\nn\\
    &=& \sum_\mathbf{k}\sum_{\zeta=\pm}\ln[1+\cosh(\beta E_\zeta)],
\eea
with the dispersion relations $E_{\pm}$ as defined in Eq.~\eqref{eq:Epm},
and the Matsubara sum performed by standard methods up to an irrelevant constant, since we are interested only in the derivatives of this quantity. Then, a careful calculation of the corresponding derivatives leads to the expression for the specific heat per unit volume for the SC phases
\bea
C_v^{SC} &=& \int\frac{d^3 k}{(2\pi)^2}\sum_{\zeta=\pm}\frac{\beta^2}{1+\cosh(\beta E_\zeta)}\Bigg\{
\bigg(E_\zeta+\beta\frac{\partial E_\zeta}{\partial\beta}\bigg)^2\nn\\
&+&\frac{\mu h_\mathbf{k}}{E_\zeta}\bigg(E_\zeta+\beta\frac{\partial E_\zeta}{\partial\beta}\bigg)+\sinh(\beta E_\zeta)\bigg[\beta\frac{\partial^2 E_\zeta}{\partial\beta^2}\nn\\
&+&\frac{\partial E_\zeta}{\partial\beta}\bigg(2-\frac{\mu h_\mathbf{k}}{E_\zeta}\bigg)\bigg]\Bigg\},
\label{eq:cvSC}
\eea
where the derivative of the energy dispersion with respect to the temperature is given in terms of the pairing gaps, as follows
\bea
\frac{\partial E_\zeta}{\partial\beta}&=&\frac{1}{2E_\zeta}\Bigg(\frac{\partial |\Delta^{\text{inter}}_\mathbf{k}|^2}{\partial\beta}
+\frac{\partial |\Delta^{\text{intra}}_\mathbf{k}|^2}{\partial\beta}\nn\\
&+&2\zeta\cos(s\phi_\mathbf{k})\frac{\partial}{\partial\beta}\big(|\Delta^{\text{inter}}_\mathbf{k}||\Delta^{\text{intra}}_\mathbf{k}|\big)\Bigg),
\eea
as well as for the second derivatives
\bea
    \frac{\partial^2 E_\zeta}{\partial\beta^2}&=&-\frac{1}{E_\zeta}\left(\frac{\partial E_\zeta}{\partial\beta}\right)^2+\frac{1}{2E_\zeta}\Bigg(\frac{\partial^2 |\Delta^{\text{inter}}_\mathbf{k}|^2}{\partial\beta^2}\\
    &+&\frac{\partial^2 |\Delta^{\text{intra}}_\mathbf{k}|^2}{\partial\beta^2}+2\zeta\cos(s\phi_\mathbf{k})\frac{\partial^2}{\partial\beta^2}\big(|\Delta^{\text{inter}}_\mathbf{k}||\Delta^{\text{intra}}_\mathbf{k}|\big)\Bigg).\nn
\eea
In the vicinity of each of the critical points, since the corresponding pairing suddenly becomes non-zero, the specific heat exhibits a discontinuity proportional to terms of the form
\begin{equation}
    \frac{\partial |\Delta^\eta|}{\partial\beta}\sim \frac{\partial}{\partial\beta}\left| T - T_c^{\eta} \right|^{1/2}
\end{equation}
due to the critical behavior of the pairing gaps.
In contrast, the metallic component of the specific heat (per unit volume) arising from the normal phase is given by the expression,
\bea
C_v^{norm} = \beta^2\int \frac{d^3 k}{(2\pi)^3}\frac{\left( \hbar v_F k - \mu  \right)^2 e^{-\beta\left(\hbar v_F k - \mu \right)}}{\left(  1 + e^{-\beta\left(\hbar v_F k - \mu \right)}\right)^2} 
\eea
that in the limit of zero chemical potential $\mu\rightarrow 0$ reduces to
\bea
C_v^{norm} &=& \frac{T^3}{2\pi^2(\hbar v_F)^3}\int_0^{\infty}dx\frac{x^4 e^{-x}}{\left( 1 + e^{-x} \right)^2}\nn\\
&=& \frac{7 \pi^2}{60}\frac{T^3}{(\hbar v_F)^3}.
\label{eq:cvNorm}
\eea
The total specific heat involving the metallic and superconducting contributions from Eq.~\eqref{eq:cvNorm} and Eq.~\eqref{eq:cvSC}, respectively, is displayed as a function of temperature in Fig.~\ref{fig:specific_heat}. The phase transitions involving each of the superconducting phases are appreciated by the discontinuities in the specific heat, arising at each of the two different critical temperatures.
\section{Phase Coexistence in the Zero Temperature Limit}
\label{Sec:Phase}
To study the possible coexistence between different phases, in this section we shall focus on the zero temperature limit. Then, in Eq.~\eqref{gap_eq_intra} and Eq.~\eqref{gap_eq_inter} the function $\mathcal{T}_\beta\to\frac{1}{2E}$. Therefore, we can compute directly the integral over $E$ as
\bea
    I_\pm=\int^{\omega_D}_{-\omega_D}\frac{dE}{\sqrt{E^2+\Delta^2_{\pm}}}=2\sinh^{-1}(\omega_D/\Delta_\pm),
\eea
where we defined $\Delta^2_\pm=|\Delta_0(\mathbf{q})|^2+|\Delta_1(\mathbf{q})|^2\pm2B_\mathbf{q}$. Then, from the mathematical identity $\sinh^{-1}(x)=\ln(x+\sqrt{x^2+1})$, and taking into account that in the weak coupling regime $|\Delta_{0,1}(\mathbf{q})|\ll\omega_D$, we obtain
\bea
    I_\pm&=&2\Bigg[\ln(2\omega_D)-\frac{1}{2}\ln(\Delta_\pm^2)\Bigg]\\
    &=&2\Bigg[\ln(2\omega_D)-\frac{1}{2}\ln\Big(|\Delta_0(\mathbf{q})|^2+|\Delta_1(\mathbf{q})|^2\pm2B_\mathbf{q}\Big)\Bigg].\nn
\label{eq:energy_integral}
\eea

In addition, the coefficients $\Delta_0^{lm}$ and $\Delta_1^{q'l'm'}$ defined in Eq.~\eqref{pairing_gap_expansion} will be determined by the symmetry of the interaction potential, as we shall illustrate in some specific examples in Section~\ref{Sec:Examples}. Making all these substitutions, we obtain the pair of coupled equations
\bea
    \frac{1}{\lambda_0}&=&\int d\Omega\Bigg\{|f_{lm}(\theta)|^2(I_++I_-)+ue^{i(m'+q'-m)\phi}\nn\\
   &\times& f_{lm}(\theta)g_{q'l'm'}(\theta)(I_+-I_-)\Bigg\}
\eea
and
\bea
    \frac{1}{\lambda_1}&=&\int d\Omega\Bigg\{|g_{q'l'm'}(\theta)|^2(I_++I_-)+u^{-1}e^{-i(m'+q'-m)\phi}\nn\\
    &\times&f_{lm}(\theta)g_{q'l'm'}(\theta)(I_+-I_-)\Bigg\},
\eea
with $u = \Delta_1^{q'l'm'}/\Delta_0^{lm}$.
We further integrate in the azimuthal angle $\phi$, thus reducing the gap equations to the expressions
\bea
    \lambda_0^{-1}&=&\int_0^\pi d\theta\sin\theta\Big\{ f_{lm}^2(\theta)\big[8\pi\ln(2\omega_D)-I_2^+-I_2^-\big]\nn\\
    &+&uf_{lm}(\theta)g_{q'l'm'}(\theta)\big[I_3^--I_3^+\big] \Big\}
\eea
and
\bea
    \lambda_1^{-1}&=&\int_0^\pi d\theta\sin\theta\Big\{ g_{q'l'm'}^2(\theta)\big[8\pi\ln(2\omega_D)-I_2^+-I_2^-\big]\nn\\
    &+&u^{-1}f_{lm}(\theta)g_{q'l'm'}(\theta)\big[I_3^--I_3^+\big]^* \Big\},
\eea
where we defined the integrals
\begin{align}
    I_2^\pm&=\int_0^{2\pi}d\phi\ln[A\pm B\cos(s\phi)],\nn\\
    I_3^\pm&=\int_0^{2\pi}d\phi\, e^{is\phi}\ln[A\pm B\cos(s\phi)],
\end{align}
and for simplicity in the notation we introduced the auxiliary variables $A=|\Delta_0(\mathbf{q})|^2+|\Delta_1(\mathbf{q})|^2$, $B=2|\Delta_0(\mathbf{q})||\Delta_1(\mathbf{q})|$, and the integer $s=m'+q'-m$. Notice that $A-B=\big(|\Delta_0(\mathbf{q})|-|\Delta_1(\mathbf{q})|\big)^2>0$, which implies the inequality $A>B$.

After a straightforward calculation, we obtain the explicit results
\begin{widetext}
\begin{equation}
    I^\pm_2=\begin{cases} 2\pi\ln\Bigg[\frac{A+\sqrt{A^2-B^2}}{2}\Bigg]=2\pi\ln\Bigg[\frac{1}{2}\Big(|\Delta_0(\mathbf{q})|^2+|\Delta_1(\mathbf{q})|^2+\big||\Delta_0(\mathbf{q})|^2-|\Delta_1(\mathbf{q})|^2\big|\Big)\Bigg] & \text{, if } s\neq0
    \\
    2\pi\ln\big[A\pm B\big]=4\pi\ln\Big[\big||\Delta_0(\mathbf{q})|\pm|\Delta_1(\mathbf{q})|\big|\Big] & \text{, if } s=0
    \end{cases},
\end{equation}
\end{widetext}
and 
\begin{widetext}
\begin{equation}
    I_3^{\pm}=\begin{cases}
        \pm2\pi\Bigg(\frac{A}{B}-\frac{\sqrt{A^2-B^2}}{B}\Bigg)=\pm\frac{\pi}{|\Delta_0(\mathbf{q})||\Delta_1(\mathbf{q})|}\Big(|\Delta_0(\mathbf{q})|^2+|\Delta_1(\mathbf{q})|^2-\big||\Delta_0(\mathbf{q})|^2-|\Delta_1(\mathbf{q})|^2\big|\Big)&\text{, if } s\neq0\\
        2\pi\ln\big[A\pm B\big]=4\pi\ln\Big[\big||\Delta_0(\mathbf{q})|\pm|\Delta_1(\mathbf{q})|\big|\Big] & \text{, if } s=0
    \end{cases}.
\end{equation}
\end{widetext}

As we can see, the gap equations adopt a different structure according to the value of the parameter $s=m'+q'-m$. For that reason, we analyze each case separately.
\begin{widetext}
\begin{equation}
    \frac{1}{\lambda_0}=\begin{cases}
    \int^\pi_0d\theta\sin(\theta)\bigg\{f^2(\theta)\ln\bigg[\frac{4\omega_D^2}{|\Delta_0^2f^2(\theta)-\Delta_1^2g^2(\theta)|}\bigg]+uf(\theta)g(\theta)\ln\bigg[\frac{|\Delta_0f(\theta)-\Delta_1g(\theta)|}{|\Delta_0f(\theta)+\Delta_1g(\theta)|}\bigg]\bigg\}, & s=0
    \\
    \int^\pi_0d\theta\sin(\theta)\bigg\{f^2(\theta)\ln\bigg[\frac{4\omega_D^2}{D_+(\theta)}\bigg]-\frac{D_-(\theta)}{\Delta_0^2}\bigg\}, & s\neq0
    \end{cases}
    \label{eq_BCS0}
\end{equation}
and
\begin{equation}
    \frac{1}{\lambda_1}=\begin{cases}
        \int^\pi_0d\theta\sin(\theta)\bigg\{g^2(\theta)\ln\bigg[\frac{4\omega_D^2}{|\Delta_0^2f^2(\theta)-\Delta_1^2g^2(\theta)|}\bigg]+\frac{f(\theta)g(\theta)}{u}\ln\bigg[\frac{|\Delta_0f(\theta)-\Delta_1g(\theta)|}{|\Delta_0f(\theta)+\Delta_1g(\theta)|}\bigg]\bigg\}, & s=0
        \\
        \int^\pi_0d\theta\sin(\theta)\bigg\{g^2(\theta)\ln\bigg[\frac{4\omega_D^2}{D_+(\theta)}\bigg]-\frac{D_-(\theta)}{\Delta_1^2}\bigg\}, & s\neq0
    \end{cases}.
    \label{eq_BCS1}
\end{equation}
\end{widetext}
Here, for notational simplicity we dropped all the indices associated with the spherical or monopole harmonics, respectively, since all the information related to them is codified in the parameter $s$ and the angular functions $f(\theta)$ and $g(\theta)$, respectively. Furthermore, we defined the auxiliary functions
\begin{equation}
    D_\pm(\theta)=\frac{1}{2}\Big(\Delta_0^2f^2(\theta)+\Delta_1^2g^2(\theta)\pm|\Delta_0^2f^2(\theta)-\Delta_1^2g^2(\theta)|\Big),
    \label{eq_Dpm}
\end{equation}
whose explicit value depends on the particular form of the angular functions within the integration domain, in addition to the value of the ratio $u = \Delta_1/\Delta_0$.

\section{Pairing Channels}
\label{Sec:pairing_channels}

To obtain the possible pairing channels admitted by our model, we take the matrix elements of the interaction in the original basis and expand them for small momenta near each Weyl node $\mathbf{K}_{\pm} = (0,0,\pm Q)$, as shown in Eq.~\eqref{eq:Vcoeff}. From these expressions, we obtain the coefficients for each conventional pairing channel as the projection over the corresponding spherical harmonic. For instance, the $s$-wave is given by the $Y_{0,0}=\frac{1}{\sqrt{4\pi}}$, so we identify the constant elements of each term, modulo a $4\pi$ factor due to normalization. Therefore we have
\begin{align}
    V^\text{intra}_{0,0}&=4\pi(V_0+3V_1)
\end{align}
and
\begin{align}
    V^\text{inter}_{0,0}&=4\pi\big(2V_0+V_1(5+\cos2Q)\big),
\end{align}
where for the inter-node case we add up all the contributions with different configurations of nodal indices.

By the same procedure, identifying the first spherical harmonics as components of the product $\mathbf{k}_\perp\cdot\mathbf{q}_\perp=\frac{4\pi}{3}kq\sum_{s=\pm1}Y_{1,s}(\Omega_\mathbf{k})Y^*_{1,s}(\Omega_\mathbf{q})$ and $k_zq_z=\frac{4\pi}{3}kqY_{1,0}(\Omega_\mathbf{k})Y^*_{1,0}(\Omega_\mathbf{q})$, we recognize the correspondign coefficients as
\begin{align}
    V^\text{intra}_{1,1}=V^\text{intra}_{1,0}=V^\text{intra}_{1,-1}=\frac{4\pi}{3}V_1
\end{align}
and
\begin{align}
    V^\text{inter}_{1,1}&=V^\text{inter}_{1,-1}=\frac{8\pi}{3}V_1,\nn\\
    V^\text{inter}_{1,0}&=\frac{4\pi}{3}V_1(1+\cos2Q).
\end{align}

The next step is to project onto the Bogoliubov basis, which diagonalizes the one-particle Hamiltonian, in order to obtain the corresponding effective pairing for each projection. For instance, the original s-channel projected over the Bogoliubov basis for the intra-node pair becomes
\begin{equation}
    \Bar{V}^\text{intra}_{\mathbf{k},\mathbf{q}}=\frac{1}{8\pi}V_{0,0}^\text{intra}[1+\cos(\theta_\mathbf{k})\cos(\theta_\mathbf{q})]e^{i(\phi_\mathbf{q}-\phi_\mathbf{k})},
\end{equation}
where the $s$-channel and the $p_z$-channel naturally appear, up to an exponential that can be removed by a suitable gauge choice. In the same manner, the projection over the inter-node pair gives
\begin{equation}
    \Bar{V}^\text{inter}_{\mathbf{k},\mathbf{q}}=\frac{1}{8\pi}V_{0,0}^\text{inter}\sin(\theta_\mathbf{k})e^{-i\phi_\mathbf{k}}\sin(\theta_\mathbf{q})e^{i\phi_\mathbf{q}},
\end{equation}
where the first monopole $\mathcal{Y}_{-1,1,0}$ is present, as a signature of the non-trivial topology of the Cooper pair for this case. Also, notice that all the possible competition between these states is determined by the condition $s = m' + q' - m\neq0$.

We can then proceed with the same reasoning for the three possible $p$-channels:
\begin{enumerate}
    \item The $p_z$-channel produces a $d$-wave pairing with couplings $\Bar{V}^\text{intra}_{2,\pm1}=\frac{1}{10}V_{1,0}^\text{intra}$, and when projected onto the inter-node pair, we obtain an infinite series in terms of monopole harmonics of the form $\mathcal{Y}_{-1,l,1}$, for all $l\geq1$, and $\mathcal{Y}_{-1,l,3}$, for $l\geq3$. Notice that in this case, again the SC states compete according to the equations with $s\neq0$, since the quantum numbers of the special functions can't satisfy $s=0$.

    \item The $p_x\pm ip_y$ channel transforms into the $d$-wave pairing with coupling $\Bar{V}^\text{intra}_{2,\pm2}=\frac{1}{5}V^\text{intra}_{1,\pm1}$, as much as the monopole-like pairing $\Bar{V}^\text{inter}_{-1,2,\pm2}=\frac{3}{10}V^\text{inter}_{1,\pm1}$. When analyzing the competition, we have to choose all the upper(lower) signs. Then, the parameter $s=-1\pm2\mp2\neq0$ again.
\end{enumerate}

The absence of competing pairings with $s=0$ can be interpreted as a manifestation of the \textit{topological repulsion} mechanism proposed in \cite{Munoz-2020}.

\section{Examples}
\label{Sec:Examples}

To illustrate the phenomenology emerging from our model, as expressed by the coupled system of BCS equations~\eqref{eq_BCS0} and \eqref{eq_BCS1}, we investigate the competition between effective pairing states arising from the microscopic potential, as discussed in the previous section. For simplicity, we choose the three pairings produced by the original $s$-wave channel: the monopole harmonic $\mathcal{Y}_{-1,1,0}$ and the two spherical harmonics $Y_{0,0}$ and $Y_{1,0}$.

\subsection{Monopole vs $s$-wave}
In this case we have to apply the condition $s\neq0$, with the angular functions $f(\theta)=1$ and $g(\theta)=\sin\theta$. Therefore, we obtain
\bea
    D_\pm(\theta)&=&\frac{1}{2}\Big(\Delta_0^2+\Delta_1^2\sin^2(\theta)\pm|\Delta_0^2-\Delta_1^2\sin^2(\theta)|\Big)\nn\\
    &\equiv&\frac{\Delta_0^2}{2}\Big(1+u^2\sin^2(\theta)\pm|1-u^2\sin^2(\theta)|\Big),
\eea
which depends upon the value of the ratio $u = \Delta_1/\Delta_0$. For instance, in the regime $u<1$ (or, equivalently, $\Delta_1<\Delta_0$) we can evaluate directly the gap equations as follows
\begin{align}
    \frac{1}{\lambda_0}&\,=2\ln\Bigg[\frac{4\omega_D^2}{\Delta_0^2}\Bigg]-\frac{4}{3}u^2,\nn \\
    \frac{1}{\lambda_1}&\,=\frac{4}{3}\Bigg(\ln\Bigg[\frac{4\omega_D^2}{\Delta_0^2}\Bigg]-1\Bigg).
\end{align}
From these equations, we obtain the condition for the phase boundary:
\begin{equation}
    \frac{2}{3}\frac{1}{\lambda_0}-\frac{1}{\lambda_1}=\frac{4}{3}\Bigg(1-\frac{2}{3}u^2\Bigg). \label{boundary_swave_1}
\end{equation}

\begin{figure}[h]
    \includegraphics[width=0.5\textwidth]{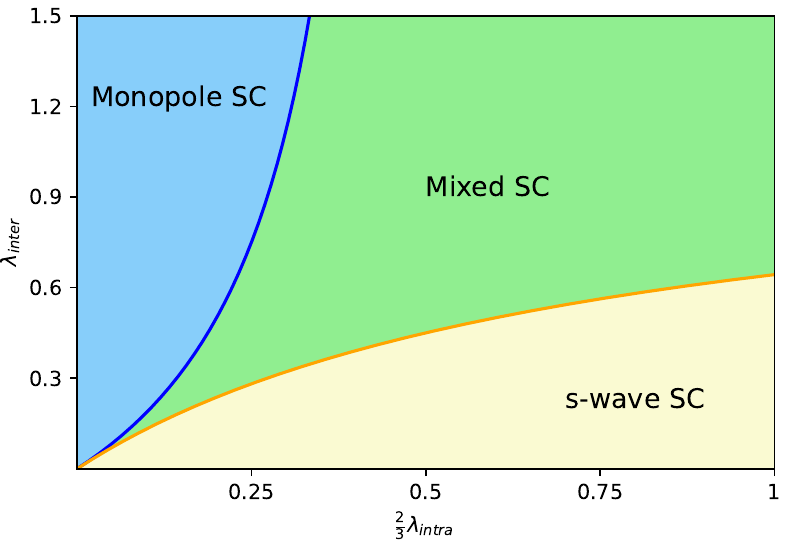}
    \caption{Phase diagram for the competition between $s$-wave SHSC vs $\mathcal{Y}_{-1,1,0}$ MSC.}
    \label{fig:phase_diagram1}
\end{figure}

For the opposite regime, $u>1$, we define the parameter $\theta_0$ such as $\sin\theta_0\equiv u^{-1}$, so the functions in Eq.~\eqref{eq_Dpm} acquire the explicit form
\begin{widetext}
\begin{equation}
    D_\pm(\theta)=\frac{\Delta_1^2}{2}\times\begin{cases}
        \sin^2(\theta_0)+\sin^2(\theta)\pm\big(\sin^2(\theta)-\sin^2(\theta_0)\big), & \theta\in[\theta_0,\pi-\theta_0] \\
        \sin^2(\theta_0)+\sin^2(\theta)\pm\big(\sin^2(\theta_0)-\sin^2(\theta)\big), & \text{elsewhere}.
    \end{cases}
\end{equation}
\end{widetext}
After a straightforward calculation, we obtain the gap equations for this regime:
\begin{widetext}
\begin{align}
    \frac{1}{\lambda_0}&\,=2\ln\Bigg[\frac{4\omega_D^2}{\Delta_0^2}\Bigg]-4\ln\Bigg[\frac{1+\cos(\theta_0)}{\sin(\theta_0)}\Bigg]+2\cos(\theta_0)-\frac{\big(\cos(3\theta_0)-9\cos(\theta_0)+8\big)}{6\sin^2(\theta_0)},\nn
    \\
    \frac{1}{\lambda_1}&\,=\frac{4}{3}\ln\Bigg[\frac{4\omega_D^2}{\Delta_0^2}\Bigg]-\frac{8}{3}\ln\Bigg[\frac{1+\cos(\theta_0)}{\sin(\theta_0)}\Bigg]+\Bigg(\frac{23}{6}-2\sin^2(\theta_0)\Bigg)\cos(\theta_0)-\frac{5}{18}\cos(3\theta_0)-\frac{4}{3}.
\end{align}
\end{widetext}
Combining them, we obtain the condition for the phase boundary, as given by the curve:
\bea
    &&\frac{2}{3}\frac{1}{\lambda_0}-\frac{1}{\lambda_1}=\cos(\theta_0)\Bigg(2\sin^2(\theta_0)+\frac{1}{\sin^2(\theta_0)}-\frac{5}{2}\Bigg)\nn\\
    &+&\frac{\cos(3\theta_0)}{9}\Bigg(\frac{5}{2}-\frac{1}{\sin^2(\theta_0)}\Bigg)-\frac{8}{9}\frac{1}{\sin^2(\theta_0)}+\frac{4}{3}.\label{boundary_swave_2}
\eea
This equation can be written in the general simpler form
\begin{equation}
    \frac{2}{3}\frac{1}{\lambda_0}-\frac{1}{\lambda_1}=f(u),
\end{equation}
where $f(u)$ has to be chosen in the two regimes according to equations (\ref{boundary_swave_1}) and (\ref{boundary_swave_2}), respectively. However, in both cases, the phase boundary curves in the phase diagram can be written as
\begin{equation}
    \lambda_1=\Bigg[\frac{2}{3}\frac{1}{\lambda_0}-f(u)\Bigg]^{-1}=\frac{3\lambda_0}{2-3\lambda_0f(u)},
\end{equation}
where the right-hand side of the equation above must be evaluated in the limits $u\to0$ for the transition from the mixed state to the $\Delta_0$ phase, and $u\to\infty$ for the transition to the $\Delta_1$ phase, respectively. The corresponding phase diagram for this case is presented in Fig.~\ref{fig:phase_diagram1}.

\subsection{Monopole vs $p_z$-wave}
In this case $s\neq0$ again, but the corresponding angular functions are $f(\theta)=\cos(\theta)$ and $g(\theta)=\sin(\theta)$, respectively. Therefore, it is convenient to make the change of variables $x=-\cos(\theta)$ in the gap equations~\eqref{eq_BCS0} and \eqref{eq_BCS1}, with the integration domain defined by $-1<x<1$. In terms of this new variable, the function in Eq.~\eqref{eq_Dpm} becomes
\begin{equation*}
    D_\pm(x)=\frac{1}{2}\Big((\Delta_0^2-\Delta_1^2)x^2+\Delta_1^2\pm|(\Delta_0^2+\Delta_1^2)x^2-\Delta_1^2|\Big).
\end{equation*}

\begin{figure}[h]
    \includegraphics[width=0.5\textwidth]{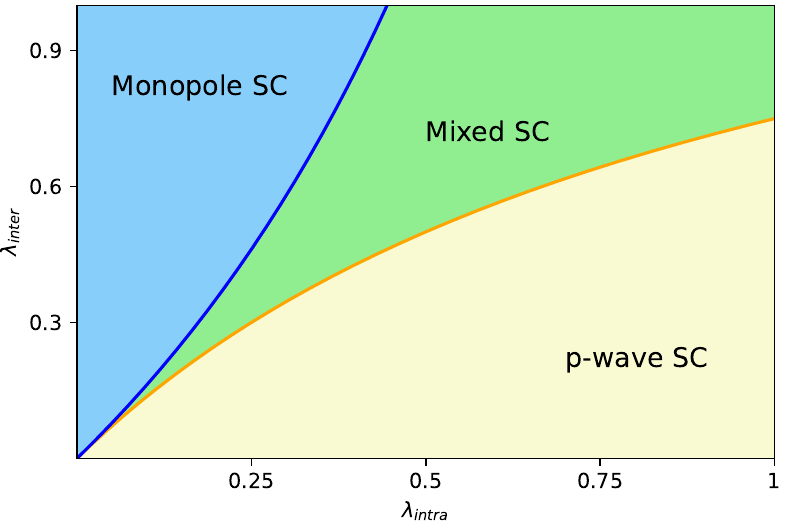}
    \caption{Phase diagram for the competition between $p_z$-wave SHSC vs $\mathcal{Y}_{-1,1,0}$ MSC.}
    \label{fig:phase_diagram2}
\end{figure}
Then, the turning points for the polynomial in the absolute value are given by the solutions to the quadratic equation
\begin{equation*}
    (\Delta_0^2+\Delta_1^2)x^2-\Delta_1^2=0,
\end{equation*}
which are given by
\bea
x_\pm=\pm\frac{\Delta_1}{\sqrt{\Delta_0^2+\Delta_1^2}}=\pm \frac{u}{\sqrt{1+u^2}}.
\eea
These solutions satisfy the condition $\left|x_{\pm}\right|\leq1$, where the equality holds only in the limit $u\to\infty$. Also, since the quadratic polynomial has positive concavity, we conclude that it takes negative values within the interval $x\in(x_-,x_+)$, and positive values elsewhere. In summary, we obtained
\begin{widetext}
\begin{equation}
    D_\pm(x)=\frac{1}{2}\times\begin{cases}
        \Delta_0^2x^2+\Delta_1^2(1-x^2)\pm\big(\Delta_1^2(1-x^2)-\Delta_0^2x^2\big) & 0<|x|<x_+
        \\
        \Delta_0^2x^2+\Delta_1^2(1-x^2)\pm\big(\Delta_0^2x^2-\Delta_1^2(1-x^2)\big) & x_+<|x|<1.
    \end{cases}
\end{equation}
\end{widetext}
Finally, since the integration domain is symmetric, and the dependence is only via the even function $x^2$, we can write the integrals as two times just the positive part of the domain. In this case, there is no distinction (from the computational point of view) between the regimes $u<1$ and $u>1$, so we can compute directly the integrals, and study the corresponding boundaries in the final result. After computing the integrals involved, we obtain the gap equations
\bea
    \frac{1}{\lambda_0}&=&\frac{2}{3}\ln\Bigg[\frac{4\omega_D^2}{\Delta_0^2}\Bigg]-\frac{2}{3}\ln\Bigg[\frac{\sqrt{1+u^2}+u}{\sqrt{1+u^2}-u}\Bigg]\nn\\
    &+&\frac{u}{\sqrt{1+u^2}}\frac{(u^2+4)}{3}-\frac{u^2}{3}+\frac{4}{9},
    \\
    \frac{1}{\lambda_1}&=&\frac{4}{3}\ln\Bigg[\frac{4\omega_D}{\Delta_0^2}\Bigg]-\frac{4}{3}\ln\Bigg[\frac{\sqrt{1+u^2}+u}{\sqrt{1+u^2}-u}\Bigg]+\frac{20}{9},\nn
\eea
which are combined in order to obtain the phase boundary condition, as defined by the curve
\begin{equation}
    \frac{1}{\lambda_0}-\frac{1}{2}\frac{1}{\lambda_1}=\frac{u}{\sqrt{1+u^2}}\frac{(u^2+4)}{3}-\frac{u^2}{3}-\frac{2}{3}.
\end{equation}
The critical behaviour predicted by this equation is depicted in figure \ref{fig:phase_diagram2}.

\subsection{Topological Repulsion}

Now we focus on the topological repulsion mechanism proposed in \cite{Munoz-2020}, by making the substitution $f_0(\theta)=f_1(\theta)\equiv f(\theta)$ in the gap equations for the case $s=0$, which leads to
\bea
    \frac{1}{\lambda_\eta}&=&\int_0^\pi d\theta\sin\theta f^2(\theta)\Bigg\{\ln\Bigg[\frac{4\omega_D^2}{f^2(\theta)|\Delta^2_\eta-\Delta^2_{\bar\eta}|}\Bigg]\nn\\
    &+&\frac{\Delta_{\bar\eta}}{\Delta_\eta}\ln\Bigg[\frac{|\Delta_\eta-\Delta_{\bar\eta}|}{|\Delta_\eta+\Delta_{\bar\eta}|}\Bigg]\Bigg\},
\label{eq:topological_repulsion}
\eea
where $\eta=0,1$ and $\bar\eta=1-\eta$. Notice that the first term is identical to the result obtained in \cite{Munoz-2020}. However, the second term gives place to a different phenomenology, beyond the topological repulsion mechanism. To see this, we take the difference between coupling constants,
\begin{align*}
    \frac{1}{\lambda_{0}}-\frac{1}{\lambda_{1}}&\,=\Bigg(\frac{\Delta_{1}}{\Delta_{0}}-\frac{\Delta_{0}}{\Delta_{1}}\Bigg)\ln\Bigg[\frac{|\Delta_{0}-\Delta_{1}|}{|\Delta_{0}+\Delta_{1}|}\Bigg]\\
    &\,=(u-u^{-1})\ln\bigg[\frac{|1-u|}{1+u}\bigg]\equiv f(u).
\end{align*}

This equation indicates that the coexistence of both phases ( $u \ne 0$ and $u<\infty$) is allowed in a continuous zone of the parameter space, since $f(u)$ is a continuous and bounded function in the whole domain $0\le u < \infty$. Therefore, we conclude that the topological repulsion mechanism does not occur in our present model. We attribute this different behaviour to the fact that our present model does not possess the asymmetry in the chemical potential considered in \cite{Munoz-2020}, which in that case leads to a non-linear term in the gap equations, thus leading to $f(u) \rightarrow 0$ in Eq.~\eqref{eq:topological_repulsion}. Thus, we conclude that asymmetry in the chemical potentials at each Weyl nodes is an \textit{essential} requirement to generate the topological repulsion mechanism.

\section{Discussion and Conclusions}
\label{Sec:Conclusions}

In this work, we investigated the existence of topologically non-trivial superconducting phases in a Weyl semimetal. For this purpose, we proposed a microscopic model for the interaction potential, involving a short-range Coulomb repulsion as well as an effective phonon-mediated attractive term. We 
further developed a self-consistent BCS-like theory for the possible pairing gaps, whose angular dependence is determined explicitly from the interactions. From this microscopic, self-consistent theory, we studied the competition between a conventional spherical harmonic (SHSC) and a monopole (MSC) superconducting phase, as a consequence of intra- and inter-nodal Cooper pairs, respectively.
We expressed our results in terms of a pair of coupled BCS equations for the pairings. In the zero temperature limit, we solved explicitly for the phase diagram and phase boundaries for different pairing channels, as directly obtained from the microscopic potential model. We found the possibility for coexistence in a Mixed SC phase, both in the s-wave and p-wave channels, respectively.
We re-examined the conditions leading to the topological repulsion mechanism proposed by us in~\cite{Munoz-2020}. Our present results allow us to conclude that, for the mechanism to take place, an asymmetry in the chemical potential at each node must be present, in combination with the selection rule $s = m' + q' - m = 0$ identified in our previous work~\cite{Munoz-2020}.

We determined the critical temperatures for both possible superconducting phases, and the critical behavior that determines the competition between the MSC and SHSC at very low temperatures. Moreover, from these finite temperature expressions, we calculated the specific heat, and showed that it exhibits discontinuities at each critical temperature, thus providing a possible experimental probe to detect the fingerprints of topological quantum criticality in Weyl semimetals. Since the existence of multiple critical temperatures is also predicted in certain conventional superconductors with broken time-reversal invariance, in order to experimentally detect the non-trivial topological effects arising from the vorticity of the resulting pairing states, magneto-transport measurements can be applied to discriminate chiral versus non-chiral pairing states. As a final remark, current ARPES measurements are very accurate at identifying the individual Weyl nodes in actual topological materials. Therefore, even though the Weyl semimetals experimentally discovered so far display several pairs of Weyl nodes at different energies in the global Fermi surface, it is in principle possible, after an ARPES characterization, to choose an appropriate gate voltage in order to set the chemical potential in close resonance with a single pair of nodes, thus fulfilling the conditions proposed in our theoretical model.

\section*{Acknowledgments}

We acknowledge financial support from ANID Fondecyt Grants No 1230440 and No 1241033.

\appendix

\section{Bogoliubov Rotation}
\label{App:Bogoliubov}
To make the change of basis $\hat{\alpha}^\dagger_a(\mathbf{k})=\xi_{a,\sigma}(\mathbf{k})\hat{\psi}^\dagger_{a,\sigma}(\mathbf{k})$, we need to find the spinor $\xi_a(\mathbf{k})$ that satisfies
\begin{equation}
    \big[v_F(k_x\hat{\sigma}^x+k_y\hat{\sigma}^y-a k_z\hat{\sigma}^z)-\mu\big]\xi_a(\mathbf{k})=\lambda^a\xi_a(\mathbf{k}),
\end{equation}
with $\lambda^a$ the corresponding energy eigenvalue. Using spherical coordinates, we can rewrite the problem as find the eigenvalues and eigenvectors of the 2x2 matrix
\begin{equation}
    M^a=\begin{pmatrix}
        -a v_Fk\cos\theta_\mathbf{k}-\mu & v_Fk\sin\theta_\mathbf{k}e^{-i\phi_\mathbf{k}} \\ v_Fk\sin\theta_\mathbf{k}e^{i\phi_\mathbf{k}} & a v_Fk\cos\theta_\mathbf{k}-\mu
    \end{pmatrix},
\end{equation}
for each possible value of $a=\pm1$.

Notice that the characteristic polynomial of $M^a$ actually does not depend on $a$, since
\bea
    &&\det [M^a-\lambda^a]=(-\mu-\lambda^a-av_Fk\cos\theta_\mathbf{k})\nn\\
    &&\times(-\mu-\lambda^a+av_Fk\cos\theta_\mathbf{k})-v_F^2k^2\sin^2\theta_\mathbf{k}
    \nn\\
    &&=(-\mu-\lambda^a)^2-(av_Fk\cos\theta_\mathbf{k})^2-v_F^2k^2\sin^2\theta_\mathbf{k}
    \nn\\
    &&=(\lambda^a+\mu)^2-v_F^2k^2,
\eea
where we used that $a^2=1$. Then, the spectrum is given by the eigenvalues $\lambda_\pm^a\equiv\lambda_\pm=\pm v_Fk-\mu$ for each node.

Since our interest is in the particle states, we just need to find the eigen-spinor associated with the eigenvalue $\lambda_+$ for each node. This is, we are looking for the kernel
\begin{equation}
    [M^a-\lambda_+]\xi_a(\mathbf{k})=0,
\end{equation}
which leads to (\ref{eq:bogoliubov_spinors}) in the main text, up to a gauge choice.

\end{document}